\documentclass[a4paper,10pt]{article}
\usepackage{amsmath}
\usepackage{amssymb}
\usepackage{epsfig}

\begin{document}

\title{``Dispersion management" for solitons in a Korteweg-de Vries system}
\author{Simon Clarke$^1$, Boris A. Malomed$^2$ and Roger Grimshaw$^3$ \\
$^1${\small Department of Mathematics and Statistics, Monash University} \\
{\small Vic. 3800, Australia} \\
$^2${\small Department of Interdisciplinary Studies, Faculty of 
Engineering,}
\\
{\small Tel Aviv University, Tel Aviv, Israel} \\
$^3${\small Department of Mathematical Sciences, Loughborough University,} 
\\
{\small Loughborough, Leicestershire LE11 3TU, UK}}
\maketitle

\begin{abstract}
The existence of ``dispersion-managed solitons'', i.e.,
stable pulsating solitary-wave solutions to the nonlinear Schr\"{o}dinger
equation with periodically modulated and sign-variable dispersion is 
now well known in nonlinear optics. Our purpose here is to investigate
whether similar structures exist for other well-known nonlinear wave models.
Hence, here we consider as a basic model the variable-coefficient 
Korteweg-de Vries equation; this has the form of a Korteweg-de Vries 
equation with a periodically varying third-order dispersion coefficient, 
that can take both positive and negative values.
More generally, this model may be extended to include
fifth-order dispersion. Such models may describe, for instance,  periodically
modulated waveguides for long gravity-capillary waves.  We develop an
analytical approximation for solitary waves in the weakly nonlinear case,
from which it is possible to obtain a reduction to a relatively simple
integral equation, which is readily solved numerically. Then, we describe
some systematic direct simulations of the full equation, which use the
soliton shape produced by the integral equation as an initial condition.
These simulations reveal regions of stable and unstable pulsating solitary
waves in the corresponding parametric space. Finally, we consider the
effects of fifth-order dispersion.
\end{abstract}

\pagebreak

\noindent \textbf{The existence of ``dispersion-managed solitons'', i.e.,
stable pulsating solitary-wave solutions to the nonlinear Schr\"{o}dinger
equation with periodically modulated dispersion, its mean value being
approximately equal to zero, is well known in nonlinear optics. However the
periodic modulation of other nonlinear integrable wave equations has received
little attention. Here we investigate the Korteweg--de Vries equation with a
periodic dispersion coefficient whose average value is small. The model
applies to long gravity-capillary waves and
natural waveguides for internal waves in a stratified ocean.
Using analytical and numerical techniques, we show that, for harmonic and
piecewise-constant profiles of the dispersion modulation, stable pulsating
solitons with small amplitudes exist. They become unstable with the increase
of the initial amplitude or mean dispersion.}

\bigskip \hrule
\bigskip

\section{Introduction and formulation of the model}

Classical models which give rise to solitons, such as the Korteweg--de Vries
(KdV) or nonlinear Schr\"{o}dinger (NLS) equations, describe uniform
nonlinear waveguides. For example, the NLS equation is a basic model to
describe light propagation in optical fibers and other guiding structures
\cite{optics}, while a well-known application of the KdV equation is the
description of internal waves in stratified fluids, propagating in
waveguides which exist naturally in the ocean, or can be created in a
laboratory \cite{internal_waves,grimshaw}. In both these models, a (bright)
soliton exists under a certain condition: in the NLS equation, the
dispersion must be anomalous \cite{optics}, while in the KdV equation, the
soliton must have a definite polarity.

The necessity to improve stability, bit rate, and other operation
characteristics for soliton streams in optical fibers, have recently
attracted a great deal of attention to the technique of ``dispersion
management'' (DM), i.e., transmission of solitons in an optical fiber which
consists of periodically alternating sections with \textit{opposite} signs
of the dispersion, so that the average dispersion of the long communication
link is nearly equal to zero \cite{DM,GT,DMgeneral}. The corresponding NLS
equation with a variable dispersion coefficient is no longer integrable by
means of the inverse scattering transform, and no exact soliton solution is
known for it. Nevertheless, very accurate numerical simulations show that,
to an extremely high accuracy, this equation gives rise to solitons with an
approximately Gaussian (rather than the ordinary $\mathrm{sech}$), \textit{%
pulsating} shape \cite{DMgeneral,GT,Anders}. The existence of \textit{%
dispersion-managed} solitons of this type is also strongly supported by
analytical results produced by the variational approximation developed for
the DM model in different forms \cite{variational,Taras}, as well as by
analysis of the model transformed into an integral equation \cite{GT,Mark}
(note that the variational approximation can also be applied directly to the
integral equation \cite{Pare}).

A remarkable property of DM solitons is that they may exist, in a stable
form, even when the average dispersion, which is not necessarily exactly
zero, but may be much smaller than the local dispersion in the alternating
fiber segments, is \emph{normal} \cite{normal,Taras} (recall that the
ordinary NLS solitons cannot exist in the case of the normal dispersion). It
is necessary to mention that, prior to the appearance of the DM models with
the above-mentioned piece-wise constant form of the dispersion profile, the
NLS equation with a harmonically (sinusoidally) modulated sign-changing
dispersion term had already been studied in detail, with a conclusion, based
on both the variational approximation and direct simulations, that stable
pulsating solitons do exist in that smoothly modulated model \cite
{sinusoidal}.

In fact, the NLS equation with the local dispersion subject to strong
periodic modulation belongs to a class of \textit{periodic heterogeneous}
systems, in which stable pulsating solitons with nontrivial properties may
be expected (in this work, we use the term ``soliton'' without assuming
integrability of the corresponding model). In nonlinear optics, other
systems belonging to this class are \textit{tandem waveguides} for optical
solitons supported by quadratic ($\chi ^{(2)}$) nonlinearity, based on
alternation of $\chi ^{(2)}$ and linear segments \cite{tandem}, \textit{%
split-step} fiber links, in which linear segments alternate with those
dominated by the Kerr ($\chi ^{(3)}$) nonlinearity \cite{splitstep}, 
\textit{%
layered} bulk media, with the $\chi ^{(3)}$ coefficient varying between the
layers so that an optical beam propagating across the layers has its power
oscillating around a critical value leading to the wave collapse \cite
{layered}, and \textit{alternate} nonlinear waveguides, composed of
periodically alternating waveguiding and \emph{anti}waveguiding segments
\cite{Gisin}. A remarkable feature, common to all these systems, despite
their very different physical nature, is robustness of the propagation
modes, and the absence of any apparent instability, even when this might
naively be expected.

The identification of this class of models, essentially based on the NLS
equation, makes it natural to ask whether other ordinary soliton-generating
equations, if subjected to a periodic modulation of the dispersion
coefficient, can give rise to the propagation of pulsating robust solitary
wave (provided that a corresponding model describes a physically meaningful
situation). The first candidate to be investigated is the Korteweg-de Vries
(KdV) equation. In fact, the KdV equation with variable coefficients is a
traditional object for the application of the perturbation theory for
solitons \cite{KN}. However, the case of periodic modulation, and the
consequent possibility of the existence of a quasi-stable pulsating soliton
in this case has not yet been studied.

We start with consideration of the propagation of weakly nonlinear long
waves in a periodically inhomogeneous waveguide, governed by the
variable-coefficient KdV equation:
\begin{equation}
u_{t}+c(\epsilon x)u_{x}+\epsilon \left[ r(\epsilon x)uu_{x}+s(\epsilon
x)u_{xxx}\right] =0(\epsilon ^{2}),  \label{okdv}
\end{equation}
where $\epsilon \ll 1$ is a basic small parameter. Equations of this general
form are commonly used to describe the propagation of solitary waves in
inhomogeneous media, for instance, surface and internal waves in stratified
fluids \cite{grimshaw}. Provided that the local nonlinear coefficient, $r$,
and phase velocity, $c$, are \emph{nonvanishing} functions of $\epsilon x$,
one can introduce a propagation coordinate $\chi $ and a temporal variable 
$%
\theta $ as follows:
\begin{equation}
\chi \equiv -\int \frac{\epsilon r}{6c^{2}}dx,\quad \theta \equiv t-\int
\frac{dx}{c}.
\end{equation}
Then Eq. (\ref{okdv}) reduces, in the same approximation at which it is
valid, to a simpler form,
\begin{equation}
u_{\chi }+6uu_{\theta }+D(\chi )u_{\theta \theta \theta }=0,  \label{KdV}
\end{equation}
where the local dispersion coefficient is now
\begin{equation}
D(\chi )=6s/\left( rc^{2}\right) .
\end{equation}
Two integral quantities, which are frequently called mass and momentum,
\begin{equation}
M=\int_{-\infty }^{+\infty }u(\theta )d\theta ,\,\,P=\int_{-\infty
}^{+\infty }u^{2}(\theta )d\theta ,  \label{MP}
\end{equation}
are exact dynamical invariants of Eq. (\ref{KdV}), i.e., $dM/d\chi =dP/d\chi
=0$. Note that $P$ may be called, more accurately, wave-action flux.

Our objective is to consider the case when $D(\chi )$ is a periodic
function, and, in particular, when it periodically changes its sign. For
instance, as it is shown in the Appendix, Eq. (\ref{KdV}) describes the
propagation of gravity-capillary waves over a periodically-varying-bottom
topography near the critical value of the Bond number; a similar derivation
can be easily carried out for interfacial-capillary waves. It is also
a possible model equation for the propagation
of internal waves and Rossby waves against the background of a slowly
varying shear flow \cite{internal_waves}. In this latter case, the wave modes
are localized near critical layers, where the long-wave speed is within the
range of the shear velocity. Generally speaking, in all these applications,
one should also take into regard the fifth-order dispersion; nevertheless,
we initially conjecture that the higher-order dispersion is negligible.

It is assumed that the variable dispersion coefficient $D(\chi )$ in Eq. (%
\ref{KdV}) can be represented in the following, quite general, form:
\begin{equation}
D=SD_{0}(\chi /T)+D_{1}(\chi ),  \label{dispersion}
\end{equation}
where $D_{0}$ is a periodic function with period $1$ (i.e., the actual
period of the first term in Eq. (\ref{dispersion}) is $T$) and amplitude 
$1$%
. The mean value of the function $D_{0}$ is \emph{exactly} equal to zero.
Following the analogy with the DM models in nonlinear optics, we will then
refer to the parameter $S$ as the map strength. The function $D_{1}$ then
represents the local average dispersion, which may also be subjected to a
long-range modulation (a situation with the average dispersion slowly
varying along the propagation distance is known for fiber-optic DM as well
\cite{Matera}). Since we are interested in the case of a sign-changing local
dispersion, it is usually assumed that $|D_{1}|\leq S/2$. In fact, without
loss of generality, one may set $S=T=1$. Indeed, it is straightforward to
see that Eq. (\ref{KdV}) is mapped into itself, but with $S=T=1$, by a
transformation to new variables,
\begin{equation}
\widetilde{u}\equiv \frac{uT}{(ST)^{1/3}},\quad \widetilde{\chi }\equiv
\frac{\chi }{T},\quad \widetilde{\theta }\equiv \frac{\theta }{(ST)^{1/3}}%
,\quad \widetilde{D}(\tau )\equiv D_{0}(\tau )+S^{-1}D_{1}(T\tau ).
\label{groups}
\end{equation}

In section 2, we develop an analytical approximation for solitary-wave
solutions to Eq. (\ref{KdV}) in the weakly nonlinear case, when the model
can be reduced to a relatively simple integral equation. Numerically found
solutions to the integral equations are also given in section 2. Results of
direct simulations of the full model are displayed in section 3,
demonstrating the existence of both stable and unstable solitons. In section
4, we briefly consider effects of the fifth-order dispersion, and the paper
is concluded by section 5. The appendix gives a short account of the
derivation of Eq. (\ref{KdV}), including the fifth-order-dispersion term,
for a particular hydrodynamic problem.

\section{Dispersion-dominated waves}

Following the approach of Ref. \cite{GT,Mark}, where the fiber-optic DM
model was transformed into an integral equation, we consider Eq. (\ref{KdV})
with the modulated dispersion in the form of Eq. (\ref{dispersion}) with $%
S=T=1$. We assume that the initial values supplementing Eq. (\ref{KdV}) are
small, viz., $f(\theta )\equiv u(\theta ,\chi =0)=O(\epsilon )$ (note that,
as it follows from Eq. (\ref{groups}), this is equivalent to assuming that 
$f=O(1)$ and $D$ varies on a fast scale). It is then natural to introduce a
slow time scale $\tau \equiv \epsilon \chi $, redefine the initial values,
so that
\begin{equation}
f(\theta )\equiv \epsilon g(\theta ),  \label{initial}
\end{equation}
and look for a solution in the form
\begin{equation}
u(\theta ,\tau )=\epsilon u^{(0)}(\theta ,\tau )+\epsilon ^{2}u^{(1)}(\theta
,\tau )+\cdots \,.  \label{epsilon}
\end{equation}
Defining the ``accumulated dispersion''
\begin{equation}
W(\chi )\equiv \int_{0}^{\chi }D_{0}(\chi )\ d\chi ,  \label{W}
\end{equation}
the zeroth-order solution to Eq. (\ref{KdV}) can be found in the form
\begin{equation}
u^{(0)}=\mathcal{F}^{-1}\left\{ \exp \left[ ik^{3}W(\chi )\right] 
\mathcal{F}%
\left\{ A(\theta ,\tau )\right\} \right\} ,  \label{zero-order}
\end{equation}
where $\mathcal{F}$ and $\mathcal{F}^{-1}$ stand for the Fourier transform
with respect to $\chi $ and its inverse. From the initial condition (\ref
{initial}), it follows that $A(\theta ,0)=g(\theta )$ in the solution (\ref
{zero-order}). Because the function $D_{0}$ is periodic with zero mean, the
functions $W(\chi )$ and $u^{(0)}$ given by Eqs. (\ref{W}) and (\ref
{zero-order}) are also periodic.

At first order in $\epsilon $, Eq. (\ref{KdV}) is reduced to
\begin{equation}
u_{\chi }^{(1)}+D_{0}u_{\theta \theta \theta }^{(1)}=-\left[ u_{\tau
}^{(0)}+6u^{(0)}u_{\theta }^{(0)}+D_{1}(\tau )u_{\theta \theta \theta 
}^{(0)}%
\right]   \label{equation1}
\end{equation}
with $u^{(1)}(\theta ,\tau =0)=0$. Defining
\begin{equation}
h(k,\chi ,\tau )\equiv \mathcal{F}\left\{ u_{\tau }^{(0)}+6u^{(0)}u_{\theta
}^{(0)}+D_{1}(\tau )u_{\theta \theta \theta }^{(0)}\right\} ,
\end{equation}
Eq. (\ref{equation1}) with the zero initial value has a solution
\begin{equation}
u^{(1)}=-\mathcal{F}^{-1}\left\{ \exp \left( ik^{3}W\right) \int_{0}^{\chi
}\exp \left[ -ik^{3}W(\sigma )\right] \,h(k,\sigma ,\tau )d\sigma \right\} .
\label{first-order}
\end{equation}
To avoid the growth of secularities in the first-order solution (\ref
{first-order}), we must impose a condition
\begin{equation}
\int_{0}^{1}\exp \left( -ik^{3}W(\chi )\right) h(k,\chi ,\tau )d\chi =0,
\end{equation}
which can be written in terms of the Fourier transform, $\hat{A}(k)$, of the
function $A(\theta )$ introduced in Eq. (\ref{zero-order}), as
\begin{equation}
\int_{0}^{1}\left\{ \hat{A}_{\tau }-iD_{0}k^{3}\hat{A}+3ik\int_{-\infty
}^{\infty }\hat{A}(\kappa )\hat{A}(k-\kappa )\exp \left[ 3ik\kappa (\kappa
-k)W(\chi )\right] \ d\kappa \right\} d\chi =0.  \label{eleven}
\end{equation}
If we introduce a function
\begin{equation}
V(\kappa )\equiv \int_{0}^{1}\exp \left[ i\kappa W(\chi )\right] \ ,
\label{kernel}
\end{equation}
then\ it follows from Eq. (\ref{eleven}) that the amplitude $A(\theta ,\tau 
)
$ satisfies an evolution equation:
\begin{equation}
A_{\tau }+3\mathcal{F}^{-1}\left\{ ik\int_{-\infty }^{\infty }V(3k\kappa
(\kappa -k))\hat{A}(\kappa )\hat{A}(k-\kappa )d\kappa \right\} +D_{0}(\tau
)A_{\theta \theta \theta }=0,  \label{weakkdv}
\end{equation}
with the initial condition that $A(\theta ,0)=g(\theta )$.

The kernel (\ref{kernel}) of the integro-differential equation 
(\ref{weakkdv}%
) can be easily calculated in the case of the piecewise-constant dispersion,
which is similar to the standard DM scheme in the fiber-optic
communications, \cite{DM,GT,DMgeneral},
\begin{equation}
D_{0}(\chi )\,=\,
\begin{cases}
-1/2\,, & 1/4<\chi <3/4, \\
+1/2, & 3/4<\chi <5/4,
\end{cases}
\label{standd}
\end{equation}
which is repeated with the period $1$. In this case, the result is
\begin{equation}
V(\kappa )=\left( 8/\kappa \right) \,\sin \left( \kappa /8\right) \,.
\label{standv}
\end{equation}
For the smooth sinusoidal modulation of the dispersion, with
\begin{equation}
D_{0}(\chi )=(1/2)\cos \left( 2\pi \chi \right) ,  \label{cosd}
\end{equation}
the kernel can be found in the form
\begin{equation}
V(\kappa )=\sum_{n=0}^{\infty }\frac{(-1)^{n}}{\left( n!\right) ^{2}}\left(
\frac{\kappa }{4}\right) ^{2n}.  \label{cosv}
\end{equation}
In general, $V(\kappa )\rightarrow 1+O(\kappa ^{2})$ 
as $\kappa \rightarrow 0$,
while for $\kappa \gg 1$, and assuming that $D_{0}(\chi )$ is sufficiently
smooth, the integral can be evaluated using the stationary-phase approximation.
The main contribution then comes from a vicinity of the point where $%
W^{\prime }(\chi )\equiv D_{0}(\chi )=0$. Assume that $D_{0}(\chi )$ has two
zeroes, $\chi _{1}$ and $\chi _{2}$, which are symmetric, such that $\chi
_{2}=1-\chi _{1}$ and $D_{0}^{\prime }(\chi _{1})=-D_{0}^{\prime }(\chi
_{2})<0$ and $W(\chi _{1})=-W(\chi _{2})$. Then, the stationary-phase method
yields
\begin{equation}
V(\kappa )\approx \sqrt{\frac{8\pi }{\kappa D_{0}^{\prime }(\chi _{2})}}\cos
\left( \kappa W(\chi _{2})+\frac{\pi }{4}\right) .
\end{equation}

Hereafter, we focus on steady-state solitary-wave solutions to Eq. (\ref
{weakkdv}), following the approach of Ref. \cite{Mark}. To this end, we
assume that the average dispersion $D_{1}$ is a positive constant, and look
for solutions in the form $A=A(\theta -S\tau )$ with a constant velocity 
$S$%
. Anticipating that $S\geq 0$, we set $S\equiv \lambda ^{2}$, so that the
Fourier transform $\hat{A}(k)$ must satisfy an equation following from Eq. 
(%
\ref{weakkdv}),
\begin{equation}
(\lambda ^{2}+D_{1}k^{2})\hat{A}=3\int_{-\infty }^{\infty }V(3k\kappa
(\kappa -k))\hat{A}(\kappa )\hat{A}(k-\kappa )d\kappa .  \label{steq}
\end{equation}
By means of the definition $\hat{A}(k)\equiv \lambda ^{2}\hat{B}(k)$, one
can combine two free parameters $D_{1}$ and $\lambda ^{2}$ into a single
one, $D_{1}/\lambda ^{2}$. Then, given the number $N$ of Fourier modes, the
following iterative scheme is used to solve the discrete truncated version
of Eq. (\ref{steq}):
\begin{align}
& \hat{B}_{k}^{(n)}=\frac{3}{\lambda ^{2}+D_{1}k^{2}}%
\sum_{l=-N/2}^{N/2}V(3kl(l-k))\hat{A}_{l}^{(n)}\hat{A}_{k-l}^{(n)},
\label{it1} \\
& \hat{A}_{k}^{(n+1)}=\frac{{\hat{P}}\{\hat{A}^{(n)}\}}{{\hat{P}}\{\hat{B}%
^{(n)}\}}\hat{B}_{k}^{(n)}\,.  \label{it2}
\end{align}
Here, the superscript is the iteration number, while the integer argument 
$k$
is written as the subscript attached to the amplitudes $\hat{B}^{(n)}$ and 
$%
\hat{A}^{(n)}$, and
\begin{equation}
{\hat{P}}\{\hat{A}\}\equiv \sum_{j=0}^{N-1}\left( \hat{A}_{j}\right) ^{2},
\end{equation}
is the momentum (see Eq. (\ref{MP})) of the discrete field $\hat{A}_{j}$. As
for similar solutions of the NLS equation in Ref. \cite{Mark}, the
shape of the steady solution converges; however, the momentum diverges, and hence
the wave must be normalized at each step of the iterative method.
\begin{figure}[tbp]
\begin{center}
\includegraphics[width=10 cm]{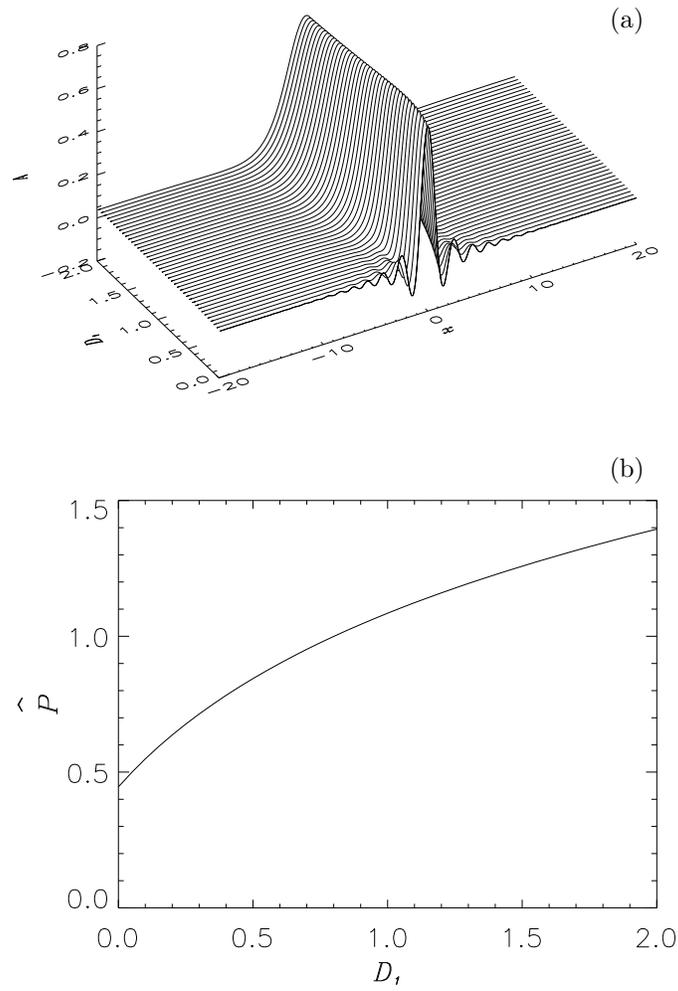} \put(-30,375){\makebox(0,0){(a)}}
\put(-30,205){\makebox(0,0){(b)}}
\end{center}
\caption{Localized solutions of Eq. (\ref{steq}), obtained for the
piecewise-constant dispersion modulation (\ref{standv}) and $\protect\lambda 
^{2}=1$. In (a) the shape of the pulse, and in (b) the variation of the
momentum $\hat{P}$ are shown as $D_{1}$ is varied.}
\label{ddsolns}
\end{figure}
\begin{figure}[tbp]
\begin{center}
\includegraphics[width=10 cm]{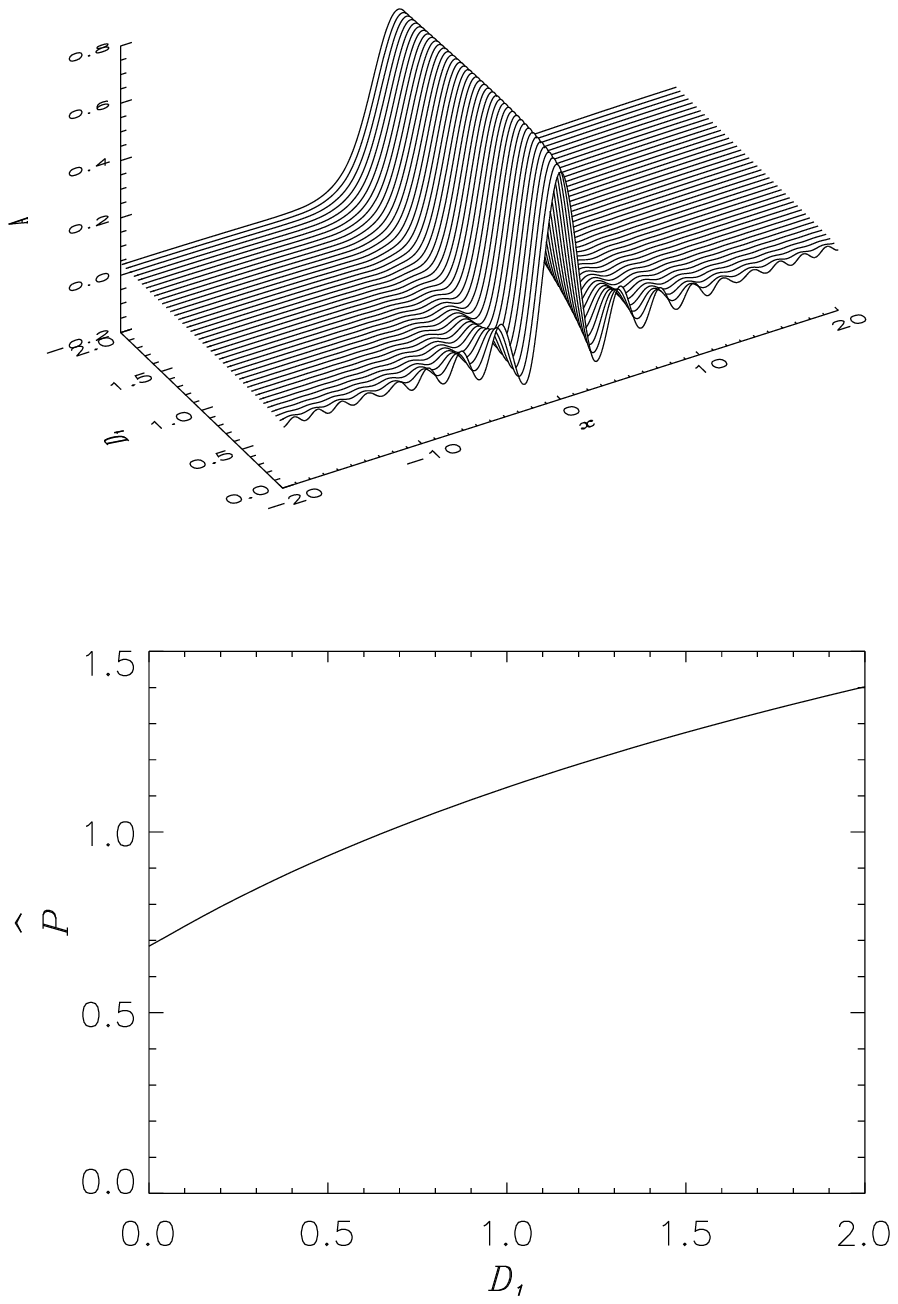} \put(-30,375){\makebox(0,0){(a)}}
\put(-30,205){\makebox(0,0){(b)}}
\end{center}
\caption{The same as in Fig. \ref{ddsolns}, but for the harmonic dispersion
modulation (\ref{cosv}).}
\label{cdsolns}
\end{figure}

In Figs. \ref{ddsolns} and \ref{cdsolns}, localized solutions obtained by
means of the numerically implemented iterative scheme (\ref{it1}), 
(\ref{it2}%
) are shown, as $D_{1}$ is varied for the two forms (\ref{standd}) and (\ref
{cosd}) of the dispersion modulation. Note that localized solutions with
positive extrema appear to be only possible for $D_{1}\geq 0$. By symmetry,
localized solutions with negative extrema and negative velocity are then
only possible for $D_{1}\leq 0$. As can be seen, a significant difference
between the waves for the two dispersion-modulation form is observed only
near $D_{1}=0$. For larger $D_{1}$, the mean dispersion dominates, and the
waves for both types of the modulation approach the ordinary $\mathrm{sech}%
^{2}$ soliton of the constant-coefficient KdV equation. For clarity, only
the domain $-20\leq x\leq 20$ is shown in these figures; in larger domains,
the oscillations apparent in Fig. \ref{cdsolns} near $D_{1}=0$ decay to 
zero.

\section{Direct numerical results}

A crucially important issue is stability of the stationary solitary waves
obtained in the previous section by means of the perturbation expansion.
This could be addressed by extending the perturbation expansion to the next
order, but this approach proves unwieldy. A more straightforward approach,
which is undertaken here, is to use solutions of Eq. (\ref{steq}) as initial
conditions for the underlying equation (\ref{KdV}), and then follow the
evolution in direct simulations. With the increase of $\epsilon $, the
nonlinearity gets stronger (see Eq. (\ref{epsilon})), leading to
stabilization or destabilization of the steady dispersion-dominated
solutions. To solve Eq. (\ref{KdV}) numerically, we used a standard method
combining pseudo-spectral techniques in $\theta $ and fourth-order
Runge--Kutta integration in $\chi $.

In this section, we consider only the case when the average dispersion 
$D_{1}
$ is constant. Further, results presented here are obtained for a slightly
smoothed version of the piecewise-constant dispersion modulation (\ref
{standd}). Results for other forms of the dispersion modulation, including
the sinusoidal form (\ref{cosd}), are very similar. Quantitative results,
such as the exact position of the stability boundary in Fig. 
\ref{stabbound}%
, see below, may be affected by the choice of the modulation form.

\begin{figure}[tbp]
\begin{center}
\includegraphics[width=10 cm]{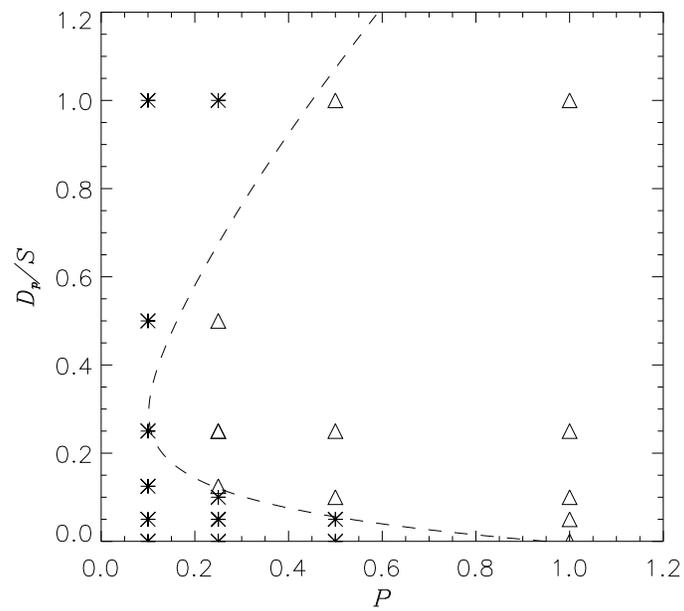}
\end{center}
\caption{The stability boundary for soliton solutions to Eq. (\ref{KdV}) for
the piecewise dispersion modulation (\ref{standd}). Stable and unstable
solutions are denoted by asterisks and triangles, respectively.}
\label{stabbound}
\end{figure}

The transformation (\ref{groups}) demonstrates that, given a particular form
of the dispersion modulation, the ratio $D_{1}/S$ and momentum $P$ uniquely
define a dispersion-dominated solution. Thus, when investigating the
stability of dispersion-dominated solutions of Eq. (\ref{KdV}), we only have
two free parameters, $D_{1}/S$ and $P$. An appropriate value of $\epsilon $
used in the previous section is then $\epsilon =P^{1/2}$. For the modulation
form (\ref{standd}), numerically found stability boundaries of the
dispersion-dominated solutions are shown in figure \ref{stabbound}.

In classifying the solutions as stable or unstable, we have used a simple
criterion that, for unstable solutions, in some region the Poincar\'{e} map
of the solution's amplitude (maximum value) will undergo sustained mean
algebraic decay in the variable $\chi $. Since we solve Eq. (\ref{KdV}) with
periodic boundary conditions in $\theta $, at a late asymptotic stage of the
evolution, the decay of the soliton will be followed by establishment of a
nearly uniform state, in which the maximum and minimum values of the field
are approximately equal and constant in $\chi $. Conversely, for stable
soliton solutions, we expect that, asymptotically, the absolute mean of the
Poincar\'{e} map of the maximum value of the field will be much larger than
its minimum-value counterpart. Strictly speaking, the location of the
stability boundary depends also on the computational domain size $L$ and
grid size $\Delta \theta $. However, we have taken care to keep $L$
sufficiently large and $\Delta \theta $ sufficiently small to minimize their
effect. The boundary also depends on the final propagation distance, $\chi
_{f}$. All the solutions shown in figure \ref{stabbound} used $\chi 
_{f}=400$%
, which was found to be sufficiently large to make it possible to conclude
whether a stable soliton was established, or the solution underwent decay.

Figure \ref{stabbound} clearly shows that there are two distinct regions of
stable solitary-wave solutions. The stability boundary corresponds to some
curve $P_{b}=P(D_{1}/S)$, with the minimum value of $P_{b}$ occurring at a
critical value $R_{c}$ of $D_{1}/S$, where for the piecewise-constant
modulation $R_{c}\approx 0.25$. Then, the stable region in $0\leq
D_{1}/S<R_{c}$ will be referred to as a ``dispersion-dominated'' one, while
the stable region in $D_{1}/S>R_{c}$ will be referred to as a ordinary
region. The numerical results suggest that the dispersion-dominated region
decreases exponentially with the increase of $D_{1}/S$, while the ordinary
region increases algebraically with the increase of $D_{1}/S$. In the
dispersion-dominated region for $D_{1}\equiv 0$, it is apparent that
localized solutions of Eq. (\ref{KdV}) are only possible up to a finite
value of the momentum. For the piecewise-constant dispersion modulation,
this critical value was found to be $P\approx 0.95$. Thus, for the
variable-dispersion KdV equation both nonlinearity and mean dispersion act
to destabilize the dispersion-dominated solitons.

The growth of the ordinary stability region with the increase of $D_{1}/S$
is easily understood. In this region, the solitary-wave solutions exist due
to balance between nonlinearity and mean dispersion, the effect of variable
dispersion is then simply to modulate the form of the wave. In this region,
solutions can be investigated using the perturbation theory for the ordinary
$\mathrm{sech}^{2}$ solitons \cite{KN}, or using methods similar to the
guiding-center technique, which were developed for solitons in optical
fibers \cite{gwp}.

\begin{figure}[tbp]
\begin{center}
\includegraphics[width=10 cm]{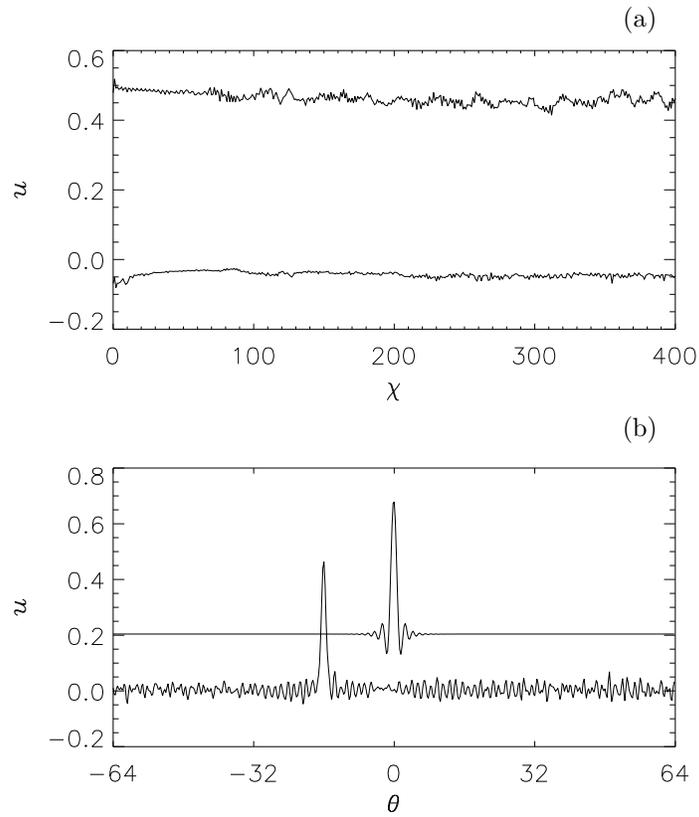} \put(-30,310){\makebox(0,0){(a)}}
\put(-30,155){\makebox(0,0){(b)}}
\end{center}
\caption{A stable dispersion-dominated solution, corresponding to the point 
$%
P=0.25$, $D_1/S = 0.05$ from Fig. \ref{stabbound}. (a) The Poincar\'e map of
the maximum and minimum values of the field, i.e., the maximum and minimum
values shown at integer values of the propagation distance
$\protect\chi$. These plots clearly show that stable nearly
stationary solitons have been found. (b) The form of the soliton at 
$\protect%
\chi = 0$ (displaced by $0.2$) and $400$.}
\label{l21}
\end{figure}

\begin{figure}[tbp]
\begin{center}
\includegraphics[width=10 cm]{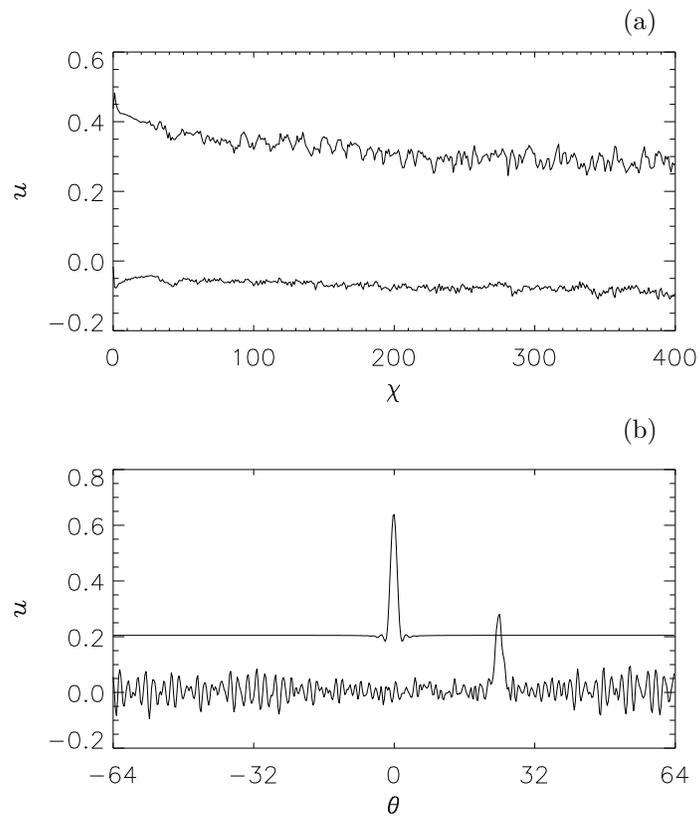} \put(-30,310){\makebox(0,0){(a)}}
\put(-30,155){\makebox(0,0){(b)}}
\end{center}
\caption{The same as in Fig. \ref{l21} for $P=0.25$ and $D_1/S = 0.25$. In
this case, the soliton is unstable.}
\label{l32}
\end{figure}

\begin{figure}[tbp]
\begin{center}
\includegraphics[width=10 cm]{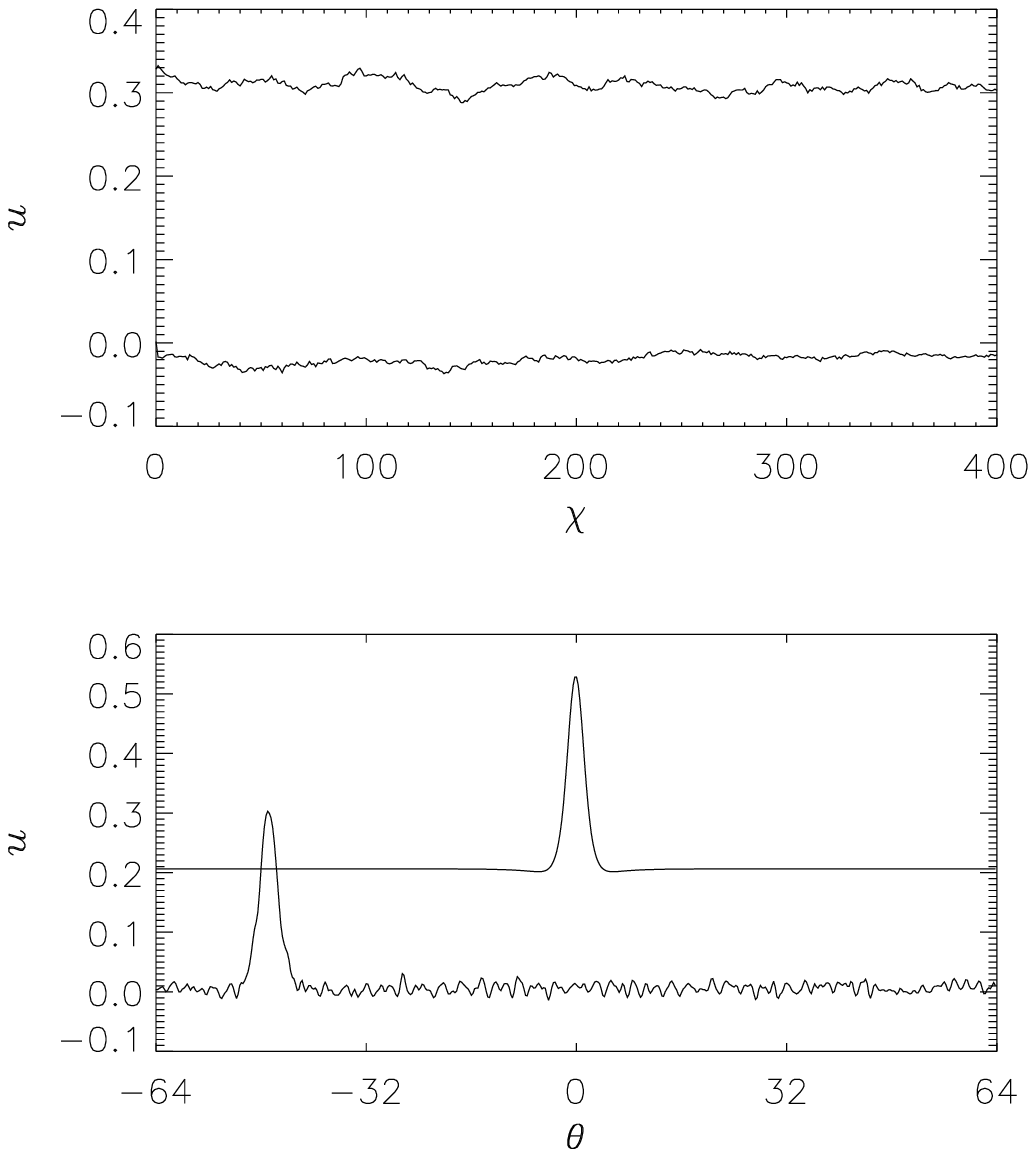} \put(-30,310){\makebox(0,0){(a)}}
\put(-30,155){\makebox(0,0){(b)}}
\end{center}
\caption{The same as in Fig. \ref{l21} for a stable soliton solution with $%
P=0.25$ and $D_1=S$ in the ``ordinary" region.}
\label{l40}
\end{figure}

In Figs. \ref{l21}, \ref{l32} and \ref{l40} examples of solutions
corresponding to particular points in Fig. \ref{stabbound} are shown. These
three soliton solutions are, respectively, stable, unstable, and stable
again. The dispersion-dominated solution in Fig. \ref{l21} demonstrates that
some momentum is lost from the localized wave to higher-mode oscillatory
waves. The subsequent interaction between the oscillatory waves and the
solitary wave causes relatively large variability in the Poincar\'{e} map of
the maximum value of the field; however, both the maximum and minimum values
eventually set down to mean values not significantly different from their
initial values. The decay of a localized wave is apparent in Fig. \ref{l32},
where the mean of both the maximum and minimum values of the field can be
seen to decrease. Eventually, at large enough propagation distance they
would be expected to become approximately equal in the absolute values and
opposite in the sign. At this stage, the mass and momentum contained
initially in the localized wave would be completely transferred to the
oscillatory waves, as is suggested by Fig. \ref{l32}(b).

For large values of the average dispersion, as is shown in Fig. \ref{l40},
which is an example of a soliton belonging to the ordinary region, it is
apparent that the solitary-wave solutions are now very similar to the $%
\mathrm{sech}^{2}$ solitons of the constant-dispersion KdV equation. Here,
only a small amount of the momentum from the initial wave is lost to the
oscillatory waves. Because of the reduced amplitude of the oscillatory wave,
the oscillations in the Poincar\'{e} map of the maximum field value are now
much slower than in Fig. \ref{l21}.

\section{Effects of fifth-order dispersion}

As was mentioned above, the fifth-order dispersion should, generally
speaking, be added to the model of the DM type with a small average
dispersion (similarly, third-order dispersion should be included in the
optical DM models \cite{TOD}). The main effect on the solitons will be to
cause them to disperse. In particular, one may expect that, for small values
of the average dispersion, the addition of the fifth-order dispersion to Eq.
(\ref{KdV}) will eventually cause stable solutions to become unstable. It is
expected that this effect is most important when the third-order dispersion
undergoes a change in sign, i.e. for $D_{1}<S/2$ in Fig. \ref{stabbound}.
For larger values of $D_{1}$, it is expected that the main effect of the
fifth-order dispersion will be to enhance the generation of oscillatory
waves \cite{skdv}. Therefore, here we briefly consider numerically the
effect of fifth-order dispersion on a stable dispersion-dominated solution.

\begin{figure}[tbp]
\begin{center}
\includegraphics[width=10cm]{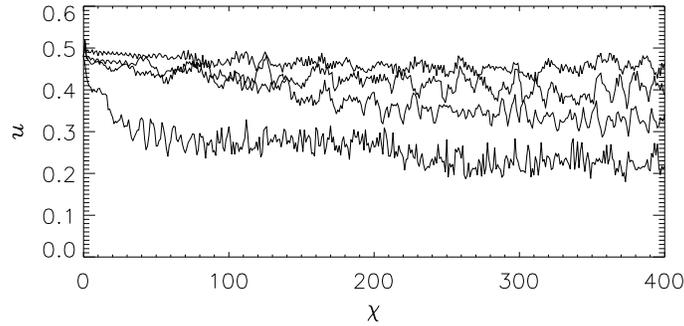}
\end{center}
\caption{The Poincar\'e map of the maximum values of the field, as obtained
from the numerical solution of Eq. (\ref{eKdV}) for $\protect\delta^2=0$, $%
10^{-3}$, $5\times10^{-3}$, and $10^{-2}$. Increasing $\protect\delta^2$
corresponds to quicker decay of the amplitude. This shows the effect of the
fifth-order dispersion on the otherwise stable dispersion-dominated solution
from Fig. \ref{l21}. }
\label{fodcomp}
\end{figure}

With the addition of fifth-order dispersion, Eq. (\ref{KdV}) becomes
\begin{equation}
u_{\chi }+6uu_{\theta }+D(\chi )u_{\theta \theta \theta }+\delta
^{2}u_{\theta \theta \theta \theta \theta }=0,  \label{eKdV}
\end{equation}
where, as shown for the case of gravity-capillary waves in the Appendix, one
can assume that the coefficient $\delta ^{2}$ is constant, in contrast to
the strongly modulated third-order-dispersion coefficient. In Fig. \ref
{fodcomp}, the effect of varying this parameter on the stable solution from
Fig. \ref{l21} is shown. These numerical solutions are obtained using a
simple extension of the pseudo-spectral Runge-Kutta method described in the
previous section. First, the case of $\delta ^{2}=10^{-4}$ was investigated,
but it is not shown in Fig. \ref{fodcomp}, as it turns out to be practically
identical to $\delta ^{2}=0$. Thus, for sufficiently small $\delta ^{2}$,
stable solitary waves are established as well as in the case $\delta 
^{2}=0$%
. However, as $\delta ^{2}$ is increased, the solitons disperse indeed,
leaving oscillatory-wave radiation. Other simulations in the
dispersion-dominated region confirm that the critical (maximum) value of $%
\delta ^{2}$, at which the solitary waves are able to form, decreases as
both $D_{1}/S$ and $P$ increase, so that the critical value is $\delta 
^{2}=0
$ on the stability boundary in Fig. \ref{stabbound}.

\section{Conclusions}

In this work we have introduced a KdV model with a periodically varying
dispersion coefficient that takes both positive and negative values. Using a
perturbative expansion based on integral equations and direct numerical
simulations, we have shown that for the case when the dispersion undergoes a
periodic change in sign, stable solitary-wave solutions are possible in a
region defined by the average dispersion and the initial momentum of the
system. Further, it has been shown that sufficiently weak fifth-order
dispersion, which, generally speaking, should be taken into regard when the
average third-order dispersion is very small, does not destroy these
solitary waves.

It should be noted that only one analytical approach (based on the integral
equation) to investigate the possibility of solitary-wave solutions has been
presented in detail in this work. We have also tried two additional
approaches. The first of these is the balance-equation method. Equation 
(\ref
{KdV}) can be derived from the Hamiltonian
\begin{equation}
H=\int_{-\infty }^{+\infty }\left[ -\frac{1}{2}D(\chi )u_{\theta 
}^{2}+u^{3}%
\right] d\theta ,  \label{H}
\end{equation}
which, unlike the mass and momentum given by Eq. (\ref{MP}), is not
conserved when the dispersion coefficient is variable. Instead, it evolves
with $\chi $ according to an immediate corollary of (\ref{KdV}),
\begin{equation}
\frac{dH}{d\chi }+\frac{1}{2}\frac{dD}{d\chi }\int_{-\infty }^{+\infty
}u_{\theta }^{2}d\theta =0.  \label{dH/dx}
\end{equation}
The balance-equation approach then demands a choice of a slowly varying
ansatz for the form of the solitary wave. Using the conservation equations
for the mass and momentum, and the evolution equation (\ref{dH/dx}) for the
Hamiltonian, evolution equations for parameters of the ansatz can be
derived. The second approach is a numerical averaging procedure, involving
successive Poincar\'{e} maps from direct numerical simulations of Eq. (\ref
{KdV}) in an attempt to generate a stable soliton solution by iterations.
However, neither of these methods have so far proved successful for the
variable-dispersion KdV equation.

The reason for solitary-wave solutions being less forthcoming from the
variable-dispersion KdV equation than for the variable-dispersion NLS
equation may have to do with the zero-dispersion limit of these two
equations. For the KdV equation, it is a singular limit with wavebreaking
occurring in finite time, whereas for the NLS equation in the
zero-dispersion limit, only the expected phase modulation takes place.
Thus, when nonlinearity dominates over the variable dispersion in the KdV
equation, as in the limit $D_{1}/S=0$ and $P\gg 1$ in Fig. \ref{stabbound}
above, one would not expect that stable solutions are possible, as the
evolution generated by the nonlinearity is much faster than that under the
action of the variable dispersion. Nonetheless, combining the results
presented here with the well-known ones for the dispersion-managed solitons
in the NLS models of long fiber-optic links consisting of alternating
segments with anomalous and normal dispersion, we conjecture that stable
pulsating solitary waves may be possible in a vast class of nonlinear wave
equations subject to strong periodic modulation of the dispersion
coefficient.

\section*{Acknowledgment}

B.A.M. appreciates the hospitality of Loughborough University (UK) and
Monash University (Clayton, Australia).

\appendix

\bigskip

\section{Zero dispersion limits of long, weakly-nonlinear waves: 
gravity-capillary waves}

A fundamental requirement for the application of Eq. (\ref{okdv}) is that 
the linear dispersion relationship between the velocity $c$ and
the wavenumber $k$ in the long wave limit can be written as
\begin{equation}
c=c_{0} -\alpha k^2 +O(k^{4}),
\end{equation}
where $c_{0}$ is non-zero, and where for some special parameter combination,
the coefficient $\alpha = 0$. In geophysical hydrodynamics, the modulation of
the coefficients of (\ref{okdv}) can then be induced by a slowly-varying
environment, e.g., flow past periodically varying topography. We briefly
describe here an example where this may occur, namely for the flow
of a long, weakly-nonlinear, gravity-capillary waves over
periodically-varying topography when the Bond number is close to 
$1/3$. A similar derivation holds for interfacial-capillary waves.

Consider then the propagation of  gravity-capillary waves
in water of depth $h$ with a horizontal length scale $L=h/\delta $ and
amplitude $a=\epsilon h$, such that $\delta ,\epsilon \ll 1$. Let $g$ be the
gravity acceleration, $\rho $ the density of the water, and $\sigma $ the
coefficient of surface tension. The linear
dispersion relationship is
\begin{equation}
c=c_{0}(1-\alpha (hk)^{2}+\beta (hk)^{4}+O(k^{6})),
\end{equation}
where
\begin{equation}
c_{0}=(gh)^{1/2},\quad \alpha =\frac{1}{2}\left( \frac{1}{3}-B\right) ,\quad
\beta =\frac{1}{2}\left( \frac{2}{15}-\frac{B}{3}-\alpha ^{2}\right)
\end{equation}
and $B=\sigma /\rho gh^{2}$ is the Bond number. Therefore, near $B=1/3$ the
leading-order dispersion coefficient is approximately zero. This occurs for
water of the depth $h\approx 1\mathrm{cm}$, in which case one should
generally also include diffusive effects. However, we will assume here 
that these are negligible; also we note that for the analogous situation
of interfacial-capillary waves, the effect of gravity is reduced and 
correspondingly the water depth may not be quite so small, 
so that in that case it is more plausible to ignore diffusive effects.

For one-dimensional waves propagating to the right, the
evolution of the free-surface displacement $\epsilon \eta $ is then governed
by the extended KdV equation \cite{skdv}
\begin{equation}
\frac{1}{c_{0}}\frac{\partial \eta }{\partial t}+\left( 1+\epsilon \frac{%
3\eta }{2h}\right) \frac{\partial \eta }{\partial x}+\delta ^{2}\alpha 
h^{2}%
\frac{\partial ^{3}\eta }{\partial x^{3}}+\delta ^{4}\beta h^{4}\frac{%
\partial ^{5}\eta }{\partial x^{5}}=O(\epsilon ^{2},\epsilon \delta
^{2},\delta ^{6}),  \label{ekdv}
\end{equation}
Now assume\ that, in
addition, the depth is a slowly varying function of $x$, such that $%
h=h(\epsilon x)$, and introduce new variables
\begin{equation}
\chi =\epsilon x,\quad \theta =g^{1/2}\left( t-\int \frac{dx}{c}\right) .
\label{newvar}
\end{equation}
Further, we put $A\equiv \eta h^{\frac{1}{4}}$ and note that in a slowly
varying environment the wave-action flux density $A^{2}$ is conserved \cite
{grimshaw,KN}. Hence, the corrected form of Eq. (\ref{ekdv}) is
\begin{equation}
\epsilon \frac{\partial A}{\partial \chi }-\epsilon 
\frac{3A}{2h^{7/4}}\frac{%
\partial A}{\partial \theta }-\delta ^{2}\alpha h^{\frac{1}{2}}\frac{%
\partial ^{3}A}{\partial \theta ^{3}}-\delta ^{4}\beta 
h^{\frac{3}{2}}\frac{%
\partial ^{5}A}{\partial x^{5}}=O(\epsilon ^{2},\epsilon \delta ^{2},\delta
^{6}).  \label{cekdv}
\end{equation}
Typically $\alpha $ is $O(1)$; then, let $\epsilon =\delta ^{2}$, which
leads to the variable-coefficient KdV equation governing the wave
propagation,
\begin{equation}
\frac{\partial A}{\partial \chi }-\frac{3A}{2h^{7/4}}\frac{\partial A}{%
\partial \theta }-\alpha h^{\frac{1}{2}}\frac{\partial ^{3}A}{\partial
\theta ^{3}}=0.  \label{vckdv}
\end{equation}
However, $B=1/3$ is a singular limit of this equation; in this limit we set 
$%
\alpha =\delta ^{2}\gamma $ and $\epsilon =\delta ^{4}$, hence Eq. (\ref
{cekdv}) becomes
\begin{equation}
\frac{\partial A}{\partial \chi }-\frac{3A}{2h^{7/4}}\frac{\partial A}{%
\partial \theta }-\gamma h^{1/2}\frac{\partial ^{3}A}{\partial \theta ^{3}}%
-\beta h^{3/2}\frac{\partial ^{5}A}{\partial \theta ^{5}}=0,
\end{equation}
where now $\beta =1/90$. Thus, if the topography and Bond number are such
that the coefficient $\alpha $ is close to zero, and can change its sign,
then a uniformly valid model equation including the fifth-order dispersion
is
\begin{equation}
\frac{\partial A}{\partial \chi }-\frac{3A}{2h^{7/4}}\frac{\partial A}{%
\partial \theta }-\alpha h^{1/2}\frac{\partial ^{3}A}{\partial \theta 
^{3}}-%
\frac{\delta ^{2}h^{3/2}}{90}\frac{\partial ^{5}A}{\partial \theta ^{5}}=0.
\label{uvckdv}
\end{equation}

\end{document}